\documentclass[lettersize,journal]{IEEEtran}
\usepackage{amsmath,amsfonts,amssymb}
\usepackage{bm}
\usepackage{cite}
\usepackage{graphicx}
\usepackage{subfigure}

\title{On the Robustness of AoA as an Authentication Feature Under Spoofing: Fundamental Limits from Misspecified Cramér–Rao Theory}
\author{Sotiris Skaperas,~\IEEEmembership{Member,~IEEE}, Arsenia Chorti,~\IEEEmembership{Senior Member,~IEEE}%
\thanks{Sotiris Skaperas and Arsenia Chorti are with the ETIS UMR 8051, CYU, ENSEA, CNRS (e-mails: {sotiris.skaperas,arsenia.chorti} @ensea.fr). The authors have been supported by the EC through the Horizon Europe/JU SNS project ROBUST-6G (Grant Agreement no. 101139068), the IPAL Project CONNECTING, the TalCyb Chair in Cybersecurity and by the French government under the France 2030 ANR program “PEPR Networks of the Future” projects. }
}

\begin{document}
\maketitle
\begin{abstract}
The robustness of angle-of-arrival (AoA) as a physical-layer authentication (PLA) feature under spoofing attacks is studied, assuming a digital uniform linear array verifier. The verifier estimates the AoA assuming a legitimate user’s single-source model, whereas the received signal is generated by a multi-antenna adversary at a different angle, leading to a model mismatch. Closed-form expressions are derived for the misspecified Cramér–Rao bound, the PLA decision threshold, the spoofing detection, false-alarm and misdetection probabilities. Simulation results validate the theoretical findings and highlight the impact of the signal-to-noise-ratio, array geometries, spoofing precoding and number of snapshots on authentication robustness.
\end{abstract}

\section{Introduction}
\IEEEPARstart{A}{ngle} of arrival (AoA)-based physical layer authentication (AoA-PLA) is an established approach for static node identification by exploiting the spatial signature of the received signal \cite{BITS,Fischer26}. In AoA-PLA, the verifier estimates the AoA from the received pilots and compares it with the enrolled AoA of the legitimate node\footnote{We note in passing that although a single AoA does not suffice to uniquely locate a legitimate node, extensions are straightforward by employing both azimuth and elevation angles in conjunction with the time of flight \cite{Fischer26}.}. Recent studies have proven that AoA is robust against spoofing (impersonation) \cite{Pham26,Pham_Globecom23}, by analyzing the mean square error between legitimate and adversarial observed signals. However, these works did not provide analytical results for the authentication accuracy, spoofing detection, false alarm and misdetection probabilities.

In this work, we close this gap; under spoofing, the verifier estimates the AoA under a nominal model, whereas the received observations are generated by a different adversarial model, corresponding to a precoded multi-antenna signal. As a result, AoA estimation under attack becomes a misspecified estimation problem. Leveraging the misspecified Cram\'er--Rao bound (MCRB) framework \cite{FortunatiSPM2017,abed2021misspecified} allows us to characterize analytically and obtain closed form expressions for authentication-relevant AoA estimation limits under spoofing.

The MCRB framework has in the past been used in wireless localization under hardware or geometry mismatch \cite{Chen2024HWI,Zheng2023MCRB}, while the authors in \cite{Usman2025} studied ghost-target spoofing in cooperative integrated sensing and communication localization over a multi-gNB network. To the best of our knowledge, this is the first work proposing to use the MCRB machinery in order to derive closed form expressions for the spoofing detection, false alarm and misdetection probabilities. The key contributions of this paper are as follows:
\begin{itemize}
\item We formulate AoA-PLA under spoofing as a misspecified estimation problem, derive the MCRB and the pseudo-true AoA for a ULA verifier and a spoofer with an arbitrary number of antennas and precoding vector.
\item We derive closed form expressions for the spoofing detection, false alarm and misdetection probabilities and identify the asymptotic limits of AoA-PLA.
\item We validate the analysis via Monte Carlo simulations and assess the impact of SNR, array size, snapshots, and spoofing geometry on authentication performance.
\end{itemize}

The remainder of the paper is organized as follows. In Section~II the assumed and spoofed signal models are presented, according to \cite{Pham26}. In Section~III the MCRB and the pseudo-true AoA are derived and the authentication hypothesis test probabilities are evaluated. In Section~IV simulation results are presented, while Section~V concludes the paper.

\section{System Model}
We consider a verifier with an $M$-element ULA, inter-element spacing $d$, wavelength $\lambda$, and wavenumber $\kappa \triangleq 2\pi d/\lambda$. We assume a narrowband far-field line-of-sight (LoS) channel and known, deterministic, unit-modulus pilots $\{p_k\}_{k=1}^{K}$ (e.g., $p_k\equiv1$), where $K$ is the number of snapshots. Under the assumed model, the channel state tinformation (CSI) $g_{\mathrm{u}}\in\mathbb{C}$ is assumed known; the unknown CSI case, requiring joint AoA and CSI estimation, is left for future work. 
\subsection{Assumed (Legitimate) Model}
We assume a single-antenna legitimate node with AoA $\theta_{\mathrm{u}} \in (-\pi/2,\pi/2)$ relative to the verifier, so that the received signal at snapshot $k$ is given by
\begin{equation}
\mathbf{x}^{\mathrm{u}}_k = g_{\mathrm{u}}\mathbf{a}(\theta_{\mathrm{u}})\,p_k + \mathbf{n}_k,\qquad k=1,\ldots,K,
\label{eq:assumed_1}
\end{equation}
where $\mathbf{n}_k \sim \mathcal{CN}(\mathbf{0},\sigma^{2}\mathbf{I}_M)$ is independent and identically distributed (i.i.d.) spatially white circularly symmetric complex Gaussian noise with variance $\sigma^2$ per antenna, and
$\mathbf{a}(\theta_{\mathrm{u}})
=\left[e^{-j\kappa m\sin\theta_{\mathrm{u}}}\right]_{m=0}^{M-1}$ is the ULA steering vector. For known $g_{\mathrm{u}}$ and $p_k \equiv 1$, we adopt the equivalent normalized model in \cite{Pham26}, so that (\ref{eq:assumed_1}) can be written as
\begin{equation}
\mathbf{x}^{\mathrm{u}}_k = \mathbf{a}(\theta_{\mathrm{u}}) + \mathbf{n}_k,\qquad k=1,\ldots,K.
\label{eq:assumed}
\end{equation}

\subsection{True (Spoofed) Model}
Following the spoofing model in \cite{Pham26}, an $L$-antenna impersonator is assumed so that the received signal is given by
\begin{equation}
{\mathbf{x}}^{\mathrm{A}}_k=\mathbf{s}+\mathbf{n}_k,\qquad k=1,\ldots,K,
\label{eq:true}
\end{equation}
where $\mathbf{n}_k\sim\mathcal{CN}(\mathbf{0},\sigma^2\mathbf{I}_M)$ has the same statistics as in the assumed model. The spoofed signal is expressed as $\mathbf{s}\triangleq\sum_{\ell=1}^{L} q_\ell\,\mathbf{a}(\theta^{\mathrm{A}}_{\ell}),$ where \(\theta^{\mathrm A}_{\ell}\in(-\pi/2,\pi/2)\) is the AoA of the \(\ell\)-th antenna element and \(q_\ell\in\mathbb{C}\) 
is the $\ell$-th precoding weight with \(q_\ell=\beta_\ell e^{j\phi_\ell}\), \(\beta_\ell\ge 0\), \(\phi_\ell\in\mathbb{R}\).

\section{Misspecified Cram\'er Rao Bound}
Let $\theta\in(-\pi/2,\pi/2)$ denote a generic candidate AoA in the assumed model. Accordingly, we define the mismatch vector as $\boldsymbol{\Delta}(\theta)\triangleq \mathbf s-\mathbf a(\theta)$. In the following, we use dot notation for derivatives with respect to $\theta$.
Let $\mathbf{X}=[\mathbf{x}_1,\ldots,\mathbf{x}_K]$ denote the observed snapshots at the verifier. Under the assumed model, the log-likelihood is given by $\ln p(\mathbf{X};\theta)
= -MK\ln(\pi\sigma^{2})
-\frac{1}{\sigma^{2}}\sum_{k=1}^{K}\|\mathbf{x}_k-\mathbf{a}(\theta)\|^{2}.$
Thus, the score function $u(\mathbf{X};\theta)\triangleq \frac{\partial}{\partial\theta}\ln p(\mathbf{X};\theta)$ is expressed as
\begin{math}
u(\mathbf{X};\theta)
= \frac{2}{\sigma^{2}}\sum_{k=1}^{K}\Re\!\left\{\dot{\mathbf{a}}(\theta)^{H}\big(\mathbf{x}_k-\mathbf{a}(\theta)\big)\right\},
\label{eq:score}
\end{math}
obtained 
using that $z+z^{*}=2\Re\{z\}$. Then, the negative second derivative is 
\begin{align}
&-\frac{\partial^{2}}{\partial\theta^{2}}\ln p(\mathbf{X};\theta)
= \sum_{k=1}^{K}\left[-\frac{\partial^{2}}{\partial\theta^{2}}\ln p(\mathbf{x}_k;\theta)\right] \nonumber\\
&= \frac{2}{\sigma^{2}}\sum_{k=1}^{K}\Re\!\left\{
\dot{\mathbf{a}}(\theta)^{H}\dot{\mathbf{a}}(\theta)
-\ddot{\mathbf{a}}(\theta)^{H}\big(\mathbf{x}_k-\mathbf{a}(\theta)\big)
\right\}.
\label{eq:se_de}
\end{align}

\subsection{Cram\'er Rao Bound}
Evaluating the expectation of the negative second derivative of the log-likelihood in \eqref{eq:se_de}  with respect to the assumed model \(p(\mathbf{X};\theta)\), so that \(\mathbb{E}[\mathbf{x}_k\mid\theta]=\mathbf{a}(\theta)\), $\forall{k}$, the second term vanishes and the Fisher information reduces to
\begin{equation}
    J(\theta) \triangleq \mathbb{E}_{\mathrm{assumed}}\left[-\frac{\partial^{2}}{\partial\theta^{2}}\ln p(\mathbf{X};\theta)\right]
    = \frac{2K}{\sigma^{2}}\|\dot{\mathbf{a}}(\theta)\|^{2}.
    \label{eq:J_def}
\end{equation}
We denote by $\Gamma(\theta)\triangleq \|\dot{\mathbf{a}}(\theta)\|^{2}$. For a ULA the explicit expression for \(\Gamma(\theta)\) is given by
\begin{equation}
    \Gamma(\theta) = \kappa^{2}\cos^{2}\theta \sum_{m=0}^{M-1} m^{2}
    = \kappa^{2}\cos^{2}\theta  \frac{(M-1)M(2M-1)}{6}.
    \label{eq:Gamma_ULA}
\end{equation}

Then, for $K$ snapshots, the CRB is given by
\begin{equation}
\mathrm{CRB}_K(\theta)
    = \frac{1}{J(\theta)}=\frac{3\sigma^{2}}{
      K\,\kappa^{2}\cos^{2}\theta\,(M-1)M(2M-1)}.
      \label{eq:CRB}
\end{equation}

\subsection{MCRB Under Model Mismatch}
The MCRB for the scalar parameter \(\theta\) is expressed as
\begin{equation}
\mathrm{MCRB}_K(\theta)=A(\theta)^{-1}B(\theta)A(\theta)^{-1},
    \label{eq:mcrb_gen}
\end{equation}
where \(A(\theta)\) and \(B(\theta)\) are computed over \(K\) i.i.d. snapshots. \(A(\theta)\) is defined as the expected observed information of the assumed log-likelihood, under the true model; using \(\mathbb{E}_{\mathrm{true}}[{\mathbf{x}}_k]=\mathbf{s}\), equivalently
\(\mathbb{E}_{\mathrm{true}}[{\mathbf{x}}_k-\mathbf{a}(\theta)]=\boldsymbol{\Delta}(\theta)\), $\forall{k}$, we obtain
\begin{align}
A(\theta)
&\triangleq
\mathbb{E}_{\mathrm{true}}\!\left[-\frac{\partial^{2}}{\partial\theta^{2}}
\ln p({\mathbf{X}};\theta)\right]
\nonumber\\
&=
\frac{2K}{\sigma^{2}}
\left(\Gamma(\theta)-\Re\!\left\{\ddot{\mathbf{a}}(\theta)^{H}\boldsymbol{\Delta}(\theta)\right\}\right)
=\frac{2K}{\sigma^{2}}D({\theta}).
\label{eq:J_true_final}
\end{align}

In the following we denote the scalar
\begin{equation}
D(\theta)\triangleq \Gamma(\theta)-\Re\!\left\{\ddot{\mathbf a}(\theta)^{H}\boldsymbol{\Delta}(\theta)\right\},
\label{eq:D_def}
\end{equation}
where \(D(\theta)\) is the mismatch dependent curvature term. \(B(\theta)\) is given by the second moment of the score function under model mismatch, expressed as
\begin{align}
B(\theta)\triangleq \mathbb{E}_{\mathrm{true}}\!\left[u({\mathbf{X}};\theta)^{2}\right]
&= \operatorname{Var}_{\mathrm{true}}\!\left[u({\mathbf{X}};\theta)\right] \nonumber\\
&\quad+\left(\mathbb{E}_{\mathrm{true}}\!\left[u({\mathbf{X}};\theta)\right]\right)^{2}.
\label{eq:K_def}
\end{align}

Since \(\mathbb{E}_{\mathrm{true}}[{\mathbf{x}}_k-\mathbf{a}(\theta)]=\boldsymbol{\Delta}(\theta)\), the mean of the score under the true model is given by
\begin{align}
    \mathbb{E}_{\mathrm{true}}[u({\mathbf{X}};\theta)]
    &= \frac{2K}{\sigma^{2}}\Re\!\left\{\dot{\mathbf{a}}(\theta)^{H}\boldsymbol{\Delta}(\theta)\right\}.
    \label{eq:score_mean}
\end{align}

We next compute the score variance under the true model. Given the snapshots are i.i.d. and
\({\mathbf{x}}_k=\mathbf{s}+\mathbf{n}_k\) with \(\mathbf{n}_k\sim\mathcal{CN}(\mathbf{0},\sigma^{2}\mathbf{I}_M)\) across \(k\),
the randomness in \(u({\mathbf{X}};\theta)=\sum_{k=1}^K u({\mathbf{x}}_k;\theta)\) is due solely to \(\{\mathbf{n}_k\}\), hence
\begin{align}
&\operatorname{Var}_{\mathrm{true}}[u({\mathbf{X}};\theta)]
    = \frac{4}{\sigma^{4}}\sum_{k=1}^{K}
       \operatorname{Var}\!\left(
       \Re\{\dot{\mathbf{a}}(\theta)^{H}\mathbf{n}\}\right) \nonumber\\ &
    = \frac{4}{\sigma^{4}}\sum_{k=1}^{K}\frac{1}{2}\,
       \mathbb{E}\!\left\{
       \left|\dot{\mathbf{a}}(\theta)^{H}\mathbf{n}\right|^{2}\right\}
    = \frac{2K}{\sigma^{2}}\Gamma(\theta),
    \label{eq:score_var}
\end{align}
since for a complex Gaussian random variable
\(z=\dot{\mathbf{a}}^{H}\mathbf{n}\), \(\operatorname{Var}(\Re\{z\})=\tfrac{1}{2}\mathbb{E}[|z|^{2}]\), and \(\mathbb{E}[|z|^{2}]=\sigma^{2}\|\dot{\mathbf{a}}(\theta)\|^{2}\).
Combining \eqref{eq:score_mean} and \eqref{eq:score_var} yields the second moment
\begin{align}
    B(\theta)
    &= \frac{2K}{\sigma^{2}}\Gamma(\theta) + \left(\frac{2K}{\sigma^{2}}\Re\{\dot{\mathbf a}(\theta)^{H}\boldsymbol{\Delta}(\theta)\}\right)^{2}.
    \label{eq:K_final}
\end{align}

We define the scalar
\begin{math}
    \eta(\theta)\triangleq
    \Re\{\dot{\mathbf{a}}(\theta)^{H}\boldsymbol{\Delta}(\theta)\},
    \label{eq:eta_def_repeat}
\end{math}
which quantifies the mismatch induced by the spoofed mean signal. Then, \eqref{eq:K_final} can be written as
\begin{equation}
    B(\theta)=\frac{2K}{\sigma^{2}}\Gamma(\theta)
    + \left(\frac{2K}{\sigma^{2}}\eta(\theta)\right)^{2}.
    \label{eq:beta_eq}
\end{equation}
Using \eqref{eq:J_true_final} and \eqref{eq:beta_eq}, \eqref{eq:mcrb_gen} can be written as
\begin{align}
\mathrm{MCRB}_K(\theta)
&=\frac{\sigma^{2}}{2K}
\frac{\Gamma(\theta)}{D(\theta)^{2}}
+\left(\frac{\eta(\theta)}{D(\theta)}\right)^2.
\label{eq:MCRB_general_final}
\end{align}

\subsection{Pseudo-True Parameter $\theta_0$}
The estimator-relevant MCRB is obtained by evaluating \eqref{eq:MCRB_general_final} at the pseudo-true parameter $\theta_0$, defined as the parameter that minimizes the Kullback--Leibler (KL) divergence between the true distribution $p_{\mathrm{true}}({\mathbf{X}})$ and the assumed likelihood $p({\mathbf{X}};\theta)$
\begin{equation}
\theta_0 \triangleq\arg\min_{\theta\in(-\frac{\pi}{2},\frac{\pi}{2})} 
D_{\mathrm{KL}}\!\left(p_{\mathrm{true}}({\mathbf{X}})\,\|\,p({\mathbf{X}};\theta)\right).
\label{eq:theta0_KL}
\end{equation}
Since $p_{\mathrm{true}}({\mathbf{X}})$ and $p({\mathbf{X}};\theta)$ are complex Gaussian with identical covariance $\sigma^2\mathbf{I}_{MK}$ and respective means $\mathbf{s}$ and $\mathbf{a}(\theta)$ replicated across snapshots, \eqref{eq:theta0_KL} is equivalent to
\begin{equation}
\theta_0=\arg\min_\theta \sum_{k=1}^{K}\|\mathbf{s}-\mathbf{a}(\theta)\|^2
=\arg\min_\theta \|\mathbf{s}-\mathbf{a}(\theta)\|^2 .
\end{equation}
At $\theta_0$, the stationarity condition and the score function imply
\begin{equation}
\mathbb{E}_{\mathrm{true}}\!\left[u({\mathbf{X}};\theta_0)\right]=0
\;\Longleftrightarrow\;
\eta(\theta_0)=0,
\label{eq:eta_theta0_zero}
\end{equation}
which follows by \eqref{eq:score}, \eqref{eq:eta_def_repeat}, $\boldsymbol{\Delta}(\theta)\triangleq \mathbf{s}-\mathbf{a}(\theta)$ and $\mathbb{E}_{\mathrm{true}} [{\mathbf{x}}_k]=\mathbf{s}$, $\forall k$. Evaluating \eqref{eq:MCRB_general_final} at $\theta_0$, the second term vanishes and
\begin{align}
\mathrm{MCRB}_{K}(\theta_0)
&=\frac{\sigma^{2}}{2K}\
\frac{\Gamma(\theta_0)}{D(\theta_0)^{2}}
=\left(\frac{\Gamma(\theta_0)}{D(\theta_0)}\right)^2\mathrm{CRB}_K(\theta_0).
\label{eq:MCRB_pseudo_true}
\end{align}

The closed-form expression of \eqref{eq:MCRB_pseudo_true} at $\theta_0$ follows from \eqref{eq:Gamma_ULA}, \eqref{eq:CRB}, and the expression of $D(\theta)$ derived in the Appendix. 
\subsection{PLA Binary Hypothesis Test}
Authentication at the verifier is formulated as a binary hypothesis test on the $K$ observed snapshots $\mathbf{X}=[\mathbf{x}_1,\ldots,\mathbf{x}_K]$, (obtained either from (\ref{eq:assumed_1}) or from (\ref{eq:true})):
\begin{align}
\mathcal{H}_0:\;& \mathbf{x}_k \sim \mathcal{CN}\!\big(\mathbf{a}(\theta_{\mathrm{u}}), \sigma^2 \mathbf{I}_M\big),
\quad k=1,\ldots,K,\ \text{(legitimate)} \notag\\
\mathcal{H}_1:\;& \mathbf{x}_k \sim \mathcal{CN}\!\bigg(\sum_{\ell=1}^{L}q_\ell\,\mathbf{a}(\theta^{\mathrm{A}}_{\ell}), \sigma^2 \mathbf{I}_M\bigg),
\text{ } k=1,\ldots,K. 
\label{eq:hyp}
\end{align}
We adopt a typical AoA verification test $T(\mathbf{X})\triangleq|\hat{\theta}-\theta_{\mathrm{u}}|$, where $\theta_{\mathrm{u}}$ is the legitimate AoA and
\begin{equation}
\hat{\theta}\triangleq\ \arg\min_{\theta\in(-\pi/2,\pi/2)}\sum_{k=1}^{K}\big\|\mathbf{x}_k-\mathbf{a}(\theta)\big\|^2
\label{estimator}
\end{equation}
is the maximum likelihood (ML) estimator under $\mathcal{H}_0$ (and the quasi-ML AoA estimator under $\mathcal{H}_1$);  this choice avoids explicit estimation of the spoofing parameters $\{\theta^{\mathrm{A}}_{\ell},q_\ell,L\}$.
The verifier rejects $\mathcal{H}_0$ if $T(\mathbf{X})>\tau$, where $\tau$ is chosen to meet a target false alarm probability $P_{FA}=\alpha$ and is expressed analytically under $\mathcal{H}_0$ via a Wald approximation. 

Under $\mathcal{H}_0$, $\{\mathbf{x}_k\}_{k=1}^{K}$ are i.i.d.\ given  $\theta_{\mathrm{u}}$. Since $\hat{\theta}$ in \eqref{estimator} is the ML estimator under $\mathcal{H}_0$, standard regularity conditions imply $\sqrt{K}\,(\hat{\theta}-\theta_{\mathrm{u}})\overset{a}{\sim}\mathcal{N}\!\big(0,\mathrm{CRB}_1(\theta_{\mathrm{u}})\big)$, $K\rightarrow\infty$; see \cite{kay1993}. Equivalently, $\hat\theta$ is asymptotically Gaussian with variance $\mathrm{CRB}_K(\theta_{\mathrm{u}})$, i.e., $\hat{\theta}\overset{a}{\sim}\mathcal{N}\!\big(\theta_{\mathrm{u}},\mathrm{CRB}_K(\theta_{\mathrm{u}})\big)$, as $K\rightarrow\infty$. Then, standardizing with $Z_0=(\hat{\theta}-\theta_{\mathrm{u}})/\sqrt{\mathrm{CRB}_K(\theta_{\mathrm{u}})}\overset{a}{\sim}\mathcal{N}(0,1)$, we obtain
\begin{equation}
P_{\mathrm{FA}}(\tau)\triangleq \Pr\!\left(|\hat{\theta}-\theta_{\mathrm{u}}|>\tau \mid \mathcal{H}_0\right)\approx2Q\!\left(\frac{\tau}{\sqrt{\mathrm{CRB}_K(\theta_{\mathrm{u}})}}\right).
\label{eq:pfa_def}
\end{equation}
where $Q(x)=1-\Phi(x)$ and $\Phi(\cdot)$ is the standard normal cdf. Hence, the threshold for false-alarm level $\alpha$ is given by
\begin{equation}
\tau(\alpha)=\sqrt{\mathrm{CRB}_K(\theta_{\mathrm{u}})}\;\Phi^{-1}\!\left(1-\frac{\alpha}{2}\right).
\label{thre}
\end{equation}
 For finite $K$, the threshold can also be calibrated numerically under $\mathcal H_0$, e.g., via Monte Carlo estimation of the $(1-\alpha)$-quantile of $|\hat{\theta}-\theta_{\mathrm{u}}|$.

Under $\mathcal{H}_1$, $\hat{\theta}$ is the quasi-ML estimator and, under standard regularity conditions for misspecified ML~\cite{FortunatiSPM2017}, satisfies
$\sqrt{K}\,(\hat{\theta}-\theta_0)\xrightarrow{d}\mathcal{N}\!\big(0,\mathrm{MCRB}_1(\theta_0)\big)$, 
equivalently $\hat{\theta}\overset{a}{\sim}\mathcal{N}\!\big(\theta_0,\mathrm{MCRB}_K(\theta_0)\big)$, as $K\to\infty$. Defining $\delta\triangleq\theta_0-\theta_{\mathrm{u}}$ and standardizing via $Z_1\triangleq(\hat{\theta}-\theta_0)/\sqrt{\mathrm{MCRB}_K(\theta_0)}\overset{a}{\sim}\mathcal{N}(0,1)$, the misdetection probability is expressed as
\begin{equation}
\begin{split}
P_{MD}(\tau)\triangleq \Pr\!\left(|\hat{\theta}-\theta_{\mathrm{u}}|\le \tau \mid \mathcal{H}_1\right)\\
\approx \Phi\!\left(\frac{\tau-\delta}{{\sqrt{\mathrm{MCRB}_K(\theta_0)}}}\right)
-\Phi\!\left(\frac{-\tau-\delta}{\sqrt{\mathrm{MCRB}_K(\theta_0)}}\right).
\end{split}
\label{eq:pmd_def}
\end{equation}
Hence, $P_{SD}\triangleq\Pr\!\left(|\hat\theta-\theta_{\mathrm{u}}|>\tau \mid \mathcal{H}_1\right)=1-P_{MD}$ denotes the spoofing detection probability. Under $\mathcal H_0$, the corresponding probability of legitimate node detection is given by $P_{D}=1-P_{FA}$. Key insights of this analysis are listed below:

\subsubsection{Asymptotic behavior of AoA-based authentication}
\begin{itemize}
\item \textit{For any fixed \(\tau>0\), \(P_{FA}(\tau)\to 0\) as \(K\to\infty\).} 
This follows from \eqref{eq:pfa_def}, $\frac{\tau}{\sqrt{\mathrm{CRB}_K(\theta_{\mathrm{u}})}}=\tau\sqrt{\frac{K}{\mathrm{CRB}_1(\theta_{\mathrm{u}})}}\xrightarrow[K\to\infty]{}\infty,$ 
and $Q(x)\xrightarrow[x\to\infty]{}0$.
\item \textit{If $|\delta|=|\theta_0-\theta_{\mathrm{u}}|>0$,  $P_{MD}(\tau(\alpha))\rightarrow0$ as $K\to\infty$}. Since $\tau(\alpha)\to 0$, for any $\varepsilon\in(0,|\delta|)$, $\tau(\alpha)\leq\varepsilon$ for all sufficiently large $K$, so $P_{MD} (\tau(\alpha))\le\Pr\!\big(|\hat{\theta}-\theta_{\mathrm{u}}|\le \varepsilon\mid \mathcal H_1\big)$. By the triangle inequality, \(\{|\hat{\theta}-\theta_{\mathrm{u}}|\le \varepsilon\}\subseteq \{|\hat{\theta}-\theta_0|\ge |\delta|-\varepsilon\}\) then
$P_{MD}(\tau(\alpha))
\le
\Pr\!\big(|\hat{\theta}-\theta_0|\ge |\delta|-\varepsilon\mid \mathcal H_1\big)
\to 0,$ since \(|\delta|-\varepsilon>0\) and \(\hat{\theta}\xrightarrow{p}\theta_0\). 
\item \textit{If $|\delta|=0$, $P_{MD} \not\to 0$ as $K\to\infty$}, since $P_{MD}(\tau(\alpha))\rightarrow{}2\Phi\!\left(
\Phi^{-1}\!\left(1-\frac{\alpha}{2}\right)
\sqrt{\frac{\mathrm{CRB}_1(\theta_{\mathrm{u}})}{\mathrm{MCRB}_1(\theta_0)}}\right)-1$.
\end{itemize}

\subsubsection{Variance sensitivity of AoA-PLA under spoofing} 
\begin{itemize}
\item \textit{If $|\delta|\leq\tau$, $P_{SD}$ is strictly increasing in estimation variance $\sigma_{\hat\theta}^{2}\triangleq\mathrm{MCRB}_K(\theta_0)$}. Differentiating $P_{SD}$ with respect to $\sigma_{\hat\theta}$ and using $\frac{d}{dx}Q(x)=-\phi(x)$, \(\phi(\cdot)\) the standard normal pdf:
$\frac{\partial P_{SD}}{\partial \sigma_{\hat{\theta}}}
=
\frac{1}{\sigma_{\hat{\theta}}^{2}}
\left[
(\tau-|\delta|)\phi\!\left(\frac{\tau-|\delta|}{\sigma_{\hat{\theta}}}\right)
+
(\tau+|\delta|)\phi\!\left(\frac{\tau+|\delta|}{\sigma_{\hat{\theta}}}\right)
\right]>0$, since $\phi(.)>0$ and $|\delta|\leq\tau$.   
\item \textit{If \(|\delta|>\tau\), \(P_{SD}\) is non-monotonic in \(\sigma_\theta\), with a unique minimizer $\sigma_\theta^{\star\,2}
={2|\delta|\tau}/
{\ln\!\big(\frac{|\delta|+\tau}{|\delta|-\tau}\big)}.$} The former yields from \(\partial P_{SD}/\partial \sigma_{\hat\theta}=0\). We rewrite $h(\sigma_{\hat\theta})\triangleq
\ln\!\left(\frac{|\delta|+\tau}{|\delta|-\tau}\right)
-\frac{2|\delta|\tau}{\sigma_{\hat\theta}^{2}}
=0$. Since \(|\delta|>\tau>0\), the function \(h(\sigma_{\hat\theta})\) is continuous on \((0,\infty)\), and
$h'(\sigma_{\hat\theta})=\frac{4|\delta|\tau}{\sigma_{\hat\theta}^{3}}>0,
\text{ } \forall \sigma_{\hat\theta}>0,$
hence \(h(\sigma_{\hat\theta})\) is strictly increasing. Moreover, $\lim_{\sigma_{\hat\theta}\to 0^{+}} h(\sigma_{\hat\theta})=-\infty$ and $\lim_{\sigma_{\hat\theta}\to \infty} h(\sigma_{\hat\theta})>0.$ Therefore, by the intermediate value theorem, there exists a solution \(\sigma_{\hat\theta}^\star>0\) to \(h(\sigma_{\hat\theta})=0\), while strict monotonicity implies that this solution is unique. Solving \(h(\sigma_{\hat\theta}^\star)=0\) yields
$\sigma_{\hat\theta}^{\star\,2}={2|\delta|\tau}/{\ln\!\left(\frac{|\delta|+\tau}{|\delta|-\tau}\right)}.$ Also, for $\sigma_{{\hat\theta}}^2\xrightarrow{}0$, $\partial P_{SD}/\partial \sigma_{\hat\theta}<0$ and for $\sigma_{\hat\theta}^2\xrightarrow{}\infty$, $\partial P_{SD}/\partial \sigma_{\hat\theta}>0$.   Hence, $\sigma_{\hat\theta}^{\star\,2}$ is the unique global minimum.

\end{itemize}

These key results confirm the findings in \cite{Pham26} with respect to the robustness of AoA against spoofing attacks, irrespective of the number of antennas at the adversarial node, as long as the spoofer is not in the same AoA as the legitimate node.

\section{Simulation Results}
We evaluate $P_{SD}$ and $P_{FA}$ versus the SNR (${1}/{\sigma^2}$), the number of verifier antennas $M$, snapshots $K$, the number of spoofer antennas $L$ and the attacker angular offset $\Delta=|\theta_{\mathrm{u}}-\theta^{\mathrm{A}}_\ell|$ for $\theta_{\mathrm{u}}=10^\circ$. For simplicity we adopt a co-linear spoofing benchmark $\theta_{\ell}^{\mathrm A}\equiv \theta_u+\Delta, \text{ } \forall \ell$, consistent with the narrowband far-field LoS assumption for a compact multi-antenna adversary and assume
$\sum_{\ell=1}^{L}|q_\ell|=1$  following the results in \cite{Pham26}. We set $d=\lambda/2$ (i.e., $\kappa=\pi$), $10^{5}$ Monte Carlo iterations and  $\alpha=10^{-3}$.

\begin{figure}[t]
\centering    \includegraphics[width=\linewidth]{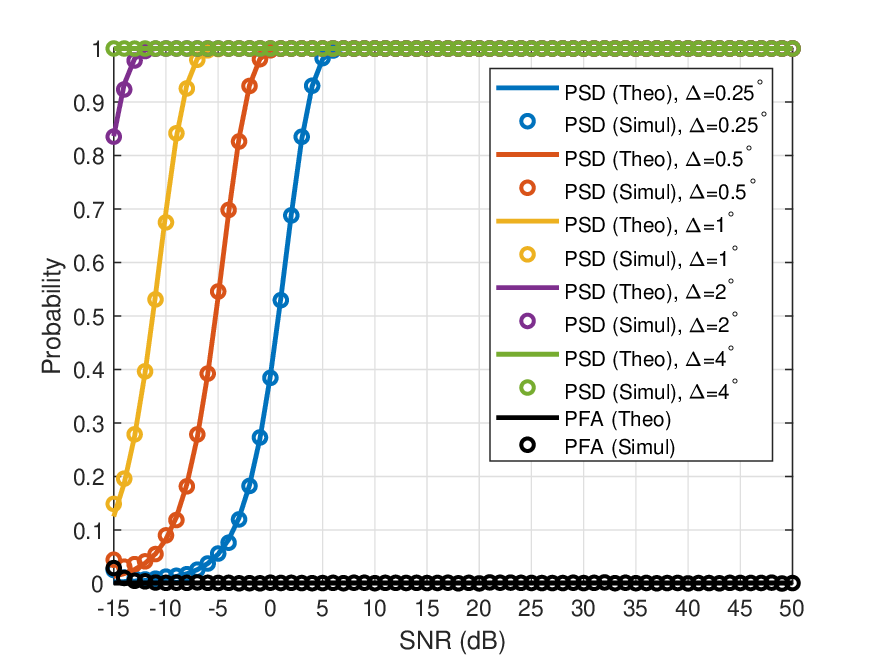}
\caption{$P_{SD}$ and $P_{FA}$ for $M=16$, $L=1$, $K=20$.}
\label{fig:SNR}
\end{figure}

In Fig.~\ref{fig:SNR}, we the effect of the SNR is studied on theoretical and simulated $P_{SD}$ and $P_{FA}$ for $\Delta=\{0.25^\circ,0.5^\circ,1^\circ,2^\circ,4^\circ\}$, with $M=16$, $L=1$, $K=20$, and SNR $\in[-15,50]$ dB. The analytical expressions closely match the corresponding simulation results. Moreover, the $P_{SD}$ increases with the SNR and $\Delta$, while the $P_{FA}$ remains close to the target level $\alpha$ across the SNR, i.e., $P_{D}\approx1$. This behavior is consistent with \eqref{eq:CRB}, \eqref{eq:MCRB_pseudo_true}, and \eqref{eq:pmd_def}: as $\sigma^2$ decreases, both $\mathrm{CRB}_K(\theta_{\mathrm{u}})$ and $\mathrm{MCRB}_K(\theta_0)$ decrease,  $\tau(\alpha)$ shrinks, and for  $\delta\neq 0$, $P_{MD}\to 0$ and $P_{SD}\to 1$ with increasing SNR. It is noteworthy that an SNR of 5 dB suffices to identify with certainty a spoofer at  an angular offest of just $\Delta=0.25^\circ$.

\begin{figure}[t]
    \centering
\includegraphics[width=\linewidth]{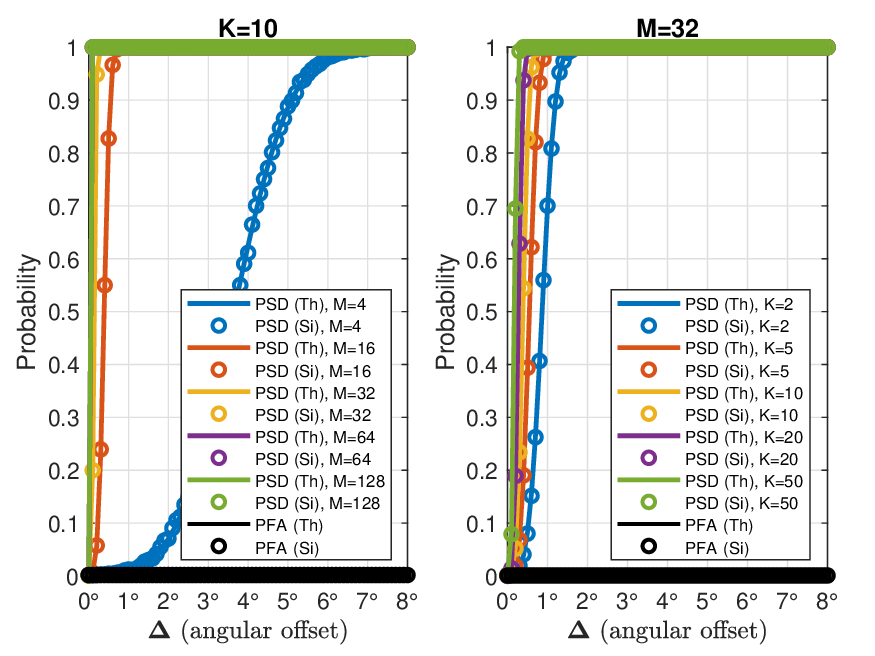}
    \caption{$P_{SD}$ and $P_{FA}$ for: a) $K=10$, and, b) $M=32$. SNR$=0$ dB.}
    \label{fig:N_M_theta}
\end{figure}

\begin{figure}[t]
    \centering
\includegraphics[width=\linewidth]{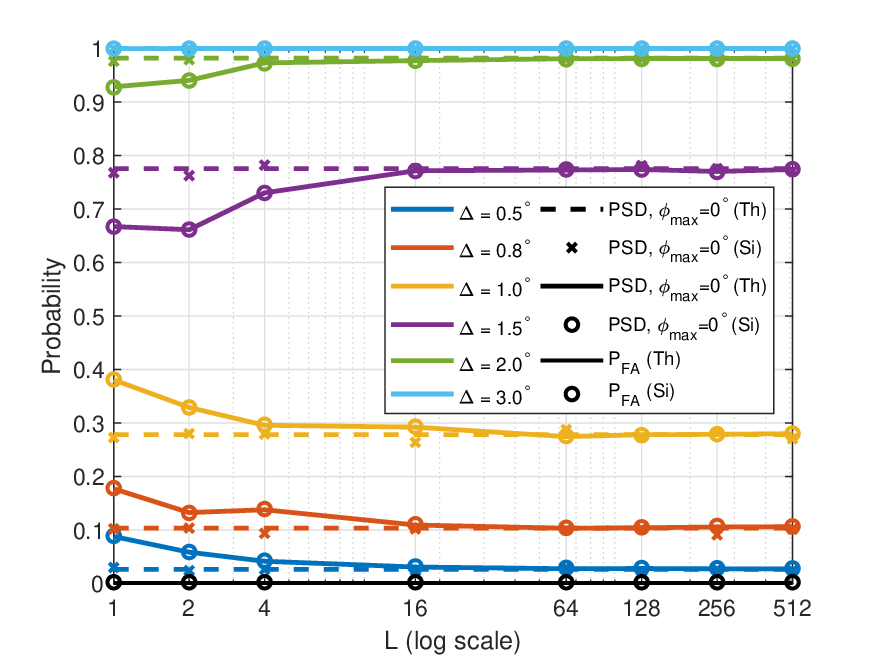}
    \caption{$P_{\mathrm{SD}}$ versus $L$ (log scale) for the equal-gain ($\phi_{\max}=0^\circ$) and the phase-impaired case ($|\phi_\ell|\leq 10^\circ$). $M=8$, $K=2$, and SNR$=5$ dB.}
    \label{fig:L_dist}
\end{figure}

In Fig.~\ref{fig:N_M_theta}, we evaluate the $P_{SD}$ and the $P_{FA}$ at SNR$=0$~dB for $\Delta\in[0^\circ,8^\circ]$ and $L=1$. In Fig.~\ref{fig:N_M_theta}(a), $K=10$ and $M=\{4,16,32,64,128\}$, while in Fig.~\ref{fig:N_M_theta}(b), $M=32$ and $K=\{2,5,10,20,50\}$. As expected, AoA-PLA is ambiguous at $\Delta=0$, whereas for $\Delta>0$, $P_{SD}$ increases rapidly with the angular offset. Larger $M$ significantly improves spoofing detection, with $M\geq16$ already achieving near-perfect performance for $\Delta<1^\circ$. Larger $K$ provides a similar but milder gain, although for $M=32$ near-perfect performance is already observed even for $K=2$. In all cases, the simulated results closely match the analytical derivations, while $P_{FA}$ remains close to the target level. These trends are consistent with \eqref{eq:CRB}, \eqref{eq:MCRB_pseudo_true}, and \eqref{eq:pmd_def}: increasing $M$ or $K$ reduces both $\mathrm{CRB}_K(\theta_{\mathrm{u}})$ and $\mathrm{MCRB}_K(\theta_0)$, which in turn shrinks $\tau(\alpha)$; thus, for any $\delta\neq0$, $P_{MD}\to0$, i.e., $P_{SD}\to1$ as the estimation variance vanishes.


Fig.~\ref{fig:L_dist} examines different precoding choices; equal-gain ($\phi_\ell=0$) vs phase-variant ($\phi_\ell\sim\mathcal U[-\phi_{\max},\phi_{\max}]$, $\phi_{\max}=10^\circ$) vs $L$, when $M=8$, $K=2$ and SNR$=5$ dB. For $\phi_\ell=0$, the $P_{\mathrm{SD}}$ is essentially invariant with respect to $L$, since the spoofed mean remains unchanged. In the phase-variant case, the effect of $L$ becomes $\Delta$-dependent through the effective coherent gain $c=\frac{1}{L}\sum_{\ell=1}^{L} e^{j\phi_\ell}.$ For small $\Delta$, the reduced coherent gain slightly increases $P_{\mathrm{SD}}$, whereas for larger $\Delta$ it slightly decreases $P_{\mathrm{SD}}$, thus favoring the attacker. This behavior reflects the variance sensitivity of the AoA test: near the acceptance region, larger estimator variance increases threshold crossing, whereas for sufficiently separated spoofed means it can reduce \(P_{\mathrm{SD}}\). As $L$ increases, the phase-variant curves move closer to the equal-gain benchmark, since the aggregate coefficient $c$ becomes increasingly concentrated around its mean. In all cases, the analytical curves closely match the simulation results, while $P_{\mathrm{FA}}$ remains near the target level. Identifying the spoofer's optimal precoding vector will be addressed in future work.

\section{Conclusions and Future Directions}
In this work a novel approach is introduced for analyzing  the impact of spoofing on AoA-PLA systems with ULA verifiers. Leveraging the MCRB framework under $\mathcal H_1$, we propose closed form expressions for the threshold $\tau(\alpha)$, the $P_{SD}$, the $P_{FA}$, and $P_{MD}$ and provide insights for the asymptotic limits of AoA-PLA. Our findings demonstrate that, in the given legitimate and attacker model and for typical SNR and array sizes, AoA is robust against spoofing and can be considered a trustworthy authentication feature. Simulated results validate the theoretical analysis and highlight the impact of the spoofing geometry and system parameters. Future work will extend the analysis to unknown CSI, multipath fading, multi-user interference and optimal spoofer precoding analysis.

\appendices
\section{Derivation of \(\eta(\theta)\) and \(D(\theta)\)}

From the definition of the ULA steering vector, let $\mathbf{\Lambda}\triangleq \operatorname{diag}(0,1,\dots,M-1)$. Then, the first- and second-order derivatives of \(\mathbf{a}(\theta)\) are
\begin{align}
\dot{\mathbf{a}}(\theta)
&= -j\kappa\cos\theta\,\mathbf{\Lambda}\mathbf{a}(\theta),\\
\ddot{\mathbf{a}}(\theta)
&= \big(j\kappa\sin\theta\,\mathbf{\Lambda}
-\kappa^{2}\cos^{2}\theta\,\mathbf{\Lambda}^{2}\big)\mathbf{a}(\theta).
\end{align}
We also define $r_\ell \triangleq e^{j\kappa(\sin\theta-\sin\theta^{\mathrm{A}}_{\ell})}, \text{for }\ell=1,\dots,L.$

From \eqref{eq:eta_def_repeat} and
$\boldsymbol{\Delta}(\theta)\triangleq \sum_{\ell=1}^{L}q_\ell\mathbf{a}(\theta^{\mathrm{A}}_{\ell})-\mathbf{a}(\theta)$,
it follows
\begin{equation}
\eta(\theta)
=
\Re\!\left\{
\sum_{\ell=1}^{L} q_{\ell}\,
\dot{\mathbf a}(\theta)^{H}\mathbf a(\theta^{\mathrm{A}}_{\ell})
-\dot{\mathbf a}(\theta)^{H}\mathbf a(\theta)
\right\}.
\label{eq:eta_split}
\end{equation}
Since \(\dot{\mathbf a}(\theta)^{H}\mathbf a(\theta)\) is purely imaginary, \eqref{eq:eta_split} reduces to
\begin{equation}
\eta(\theta)
=
\Re\!\left\{
\sum_{\ell=1}^{L} q_{\ell}\,
\dot{\mathbf a}(\theta)^{H}\mathbf a(\theta^{\mathrm{A}}_{\ell})
\right\}.
\label{eq:eta_no_self}
\end{equation}

For each attacker angle \(\theta^{\mathrm{A}}_{\ell}\),
\begin{align}
\dot{\mathbf a}(\theta)^{H}\mathbf a(\theta^{\mathrm{A}}_{\ell})
&= j\kappa\cos\theta
\sum_{m=0}^{M-1}
m\,e^{j\kappa m(\sin\theta-\sin\theta^{\mathrm{A}}_{\ell})} \nonumber\\
&= j\kappa\cos\theta\,S_1(r_\ell),
\end{align}
where $S_1(r_\ell)\triangleq \sum_{m=0}^{M-1} m r_\ell^m$ is the first weighted geometric sum. For \(r_\ell\neq 1\),
\begin{equation}
S_1(r_\ell)
=
\frac{r_\ell\big(1-Mr_\ell^{M-1}+(M-1)r_\ell^{M}\big)}{(1-r_\ell)^2},
\label{eq:geom_sum}
\end{equation}
whereas for \(r_\ell=1\), i.e., \(\theta=\theta^{\mathrm{A}}_{\ell}\),
$S_1(1)=\frac{M(M-1)}{2}$.

Substituting into \eqref{eq:eta_no_self} gives
\begin{equation}
\eta(\theta)
=
\Re\!\left\{
j\kappa\cos\theta
\sum_{\ell=1}^{L} q_{\ell}\,S_1(r_{\ell})
\right\}.
\end{equation}
Using \(\Re\{jz\}=-\Im\{z\}\), we obtain
\begin{equation}
\eta(\theta)
=
-\kappa\cos\theta\;
\Im\!\left\{
\sum_{\ell=1}^{L} q_{\ell}\,S_1(r_{\ell})
\right\}.
\label{eq:eta_final}
\end{equation}

Next, since \(\|\mathbf a(\theta)\|^2=\mathbf a(\theta)^H\mathbf a(\theta)=M\) is constant, it follows that
$
\Re\{\ddot{\mathbf a}(\theta)^H\mathbf a(\theta)\}=-\|\dot{\mathbf a}(\theta)\|^2=-\Gamma(\theta)$. Then, from \eqref{eq:D_def},
$D(\theta)
=
-\Re\!\left\{\ddot{\mathbf a}(\theta)^{H}\mathbf{s}\right\}$.

Define $\nu_\ell(\theta)\triangleq\kappa(\sin\theta-\sin\theta^{\mathrm{A}}_{\ell}).$ Then,
\begin{align}
\ddot{\mathbf a}(\theta)^{H}\mathbf a(\theta^{\mathrm{A}}_{\ell})
&=
\sum_{m=0}^{M-1}
\Big(-j\kappa m\sin\theta-\kappa^{2}m^{2}\cos^{2}\theta\Big)
e^{j m \nu_{\ell}(\theta)} \nonumber\\
&=
-j\kappa\sin\theta\,S_{1}(r_{\ell})
-\kappa^{2}\cos^{2}\theta\,S_{2}(r_{\ell}),
\end{align}
where $S_{2}(r_{\ell})\triangleq \sum_{m=0}^{M-1} m^2 r_\ell^m$ is the second weighted geometric sum. For \(r_\ell\neq 1\),
\begin{align}
S_{2}(r_\ell)
&=\frac{r_\ell}{(1-r_\ell)^{3}}
\Big(1+r_\ell-M^{2}r_\ell^{M-1} \nonumber \\ &+\big(2M^{2}-2M-1\big)r_\ell^{M} -(M-1)^{2}r_\ell^{M+1}\Big),
\end{align}
whereas for \(r_\ell=1\),
$S_2(1)=\frac{(M-1)M(2M-1)}{6}$.
Therefore,
\begin{equation}
D(\theta)=
\Re\!\left\{
j\kappa\sin\theta \sum_{\ell=1}^{L} q_{\ell}\,S_{1}(r_{\ell})
+\kappa^{2}\cos^{2}\theta \sum_{\ell=1}^{L} q_{\ell}\,S_{2}(r_{\ell})
\right\},
\end{equation}
or, equivalently,
\begin{align}
D(\theta)
&= -\kappa\sin\theta\;\Im\!\left\{\sum_{\ell=1}^{L} q_{\ell}\,S_{1}(r_{\ell})\right\} 
+ \kappa^{2}\cos^{2}\theta\;\Re\!\left\{\sum_{\ell=1}^{L} q_{\ell}\,S_{2}(r_{\ell})\right\}.
\label{d_final}\nonumber
\end{align}

\bibliographystyle{IEEEtran}
\bibliography{references}

@ARTICLE{Fischer26,
  author={Fischer, Georg K. J. and Schaechtle, Thomas and Gabbrielli, Andrea and Bordoy, Joan and Häring, Ivo and Höflinger, Fabian and Rupitsch, Stefan J.},
  journal={IEEE Commun. Surveys and Tut.}, 
  title={A Systematic Survey and Comparative Analysis of Angular-Based Indoor Localization and Positioning Technologies}, 
  year={2026},
  volume={28},
  number={},
  pages={3830-3869},
  doi={10.1109/COMST.2025.3567765}}

@ARTICLE{Pham26,
  author={Pham, Thuy M. and Senigagliesi, Linda and Baldi, Marco and Schaefer, Rafael F. and Fettweis, Gerhard P. and Chorti, Arsenia},
  journal={IEEE Trans.  Inf. Forens. Security}, 
  title={Leveraging Angle of Arrival Estimation against Impersonation Attacks in Physical Layer Authentication}, 
  year={2026},
  volume={},
  number={},
  pages={1-1},
  doi={10.1109/TIFS.2026.3675885}}

@ARTICLE{Usman2025,
  author  = {Ali, Usman and Blefari-Melazzi, Nicola and Bartoletti, Stefania},
  journal = {IEEE Wireless Commun. Lett.},
  title   = {Cooperative {ISAC} under Spoofing Attacks},
  year    = {2025},
  volume  = {14},
  number  = {9},
  pages   = {2683--2687}
}

@INPROCEEDINGS{Pham_Globecom23,
  author    = {Pham, Thuy M. and Senigagliesi, Linda and Baldi, Marco and Fettweis, Gerhard P. and Chorti, Arsenia},
  booktitle = {Proc. IEEE GLOBECOM},
  title     = {Machine Learning-Based Robust Physical Layer Authentication Using Angle-of-Arrival Estimation},
  year      = {2023},
  pages     = {13--18}
}

@ARTICLE{abed2021misspecified,
  author  = {Le Trung Thanh and Abed-Meraim, Karim and Linh-Trung, Nguyen},
  journal = {IEEE Trans. Signal Process.},
  title   = {Misspecified {C}ram{\'e}r--{R}ao Bounds for Blind Channel Estimation Under Channel Order Misspecification},
  year    = {2021},
  volume  = {69},
  pages   = {5372--5385}
}

@ARTICLE{BITS,
  author  = {Mitev, Miroslav and Pham, Thuy M. and Chorti, Arsenia and Barreto, Andr{\'e} Noll and Fettweis, Gerhard},
  journal = {IEEE BITS Inf. Theory Mag.},
  title   = {Physical Layer Security---From Theory to Practice},
  year    = {2023},
  volume  = {3},
  number  = {2},
  pages   = {67--79}
}

@INPROCEEDINGS{Zheng2023MCRB,
  author    = {Zheng, Pinjun and Chen, Hui and Ballal, Tarig and Wymeersch, Henk and Al-Naffouri, Tareq Y.},
  booktitle = {Proc. IEEE ICASSP},
  title     = {Misspecified Cram{\'e}r--Rao Bound of {RIS}-Aided Localization Under Geometry Mismatch},
  year      = {2023},
  pages     = {1--5}
}

@ARTICLE{FortunatiSPM2017,
  author  = {Fortunati, S. and Gini, F. and Greco, M. S. and Richmond, C. D.},
  title   = {Performance Bounds for Parameter Estimation Under Misspecified Models: Fundamental Findings and Applications},
  journal = {IEEE Signal Process. Mag.},
  volume  = {34},
  number  = {6},
  pages   = {142--157},
  year    = {2017},
  month   = nov,
  doi     = {10.1109/MSP.2017.2738017}
}

@ARTICLE{Chen2024HWI,
  author  = {Chen, Hui and Keskin, Musa Furkan and Aghdam, Sina Rezaei and Kim, Hyowon and Lindberg, Simon and Wolfgang, Andreas and Abrudan, Traian E. and Eriksson, Thomas and Wymeersch, Henk},
  journal = {IEEE Trans. Wireless Commun.},
  title   = {Modeling and Analysis of {OFDM}-Based {5G}/{6G} Localization Under Hardware Impairments},
  year    = {2024},
  volume  = {23},
  number  = {7},
  pages   = {7319--7333},
  doi     = {10.1109/TWC.2023.3339523}
}

@book{Kay1993,
  author    = {Steven M. Kay},
  title     = {Fundamentals of Statistical Signal Processing, Volume I: Estimation Theory},
  address   = {Philadelphia, PA},
  publisher = {Prentice Hall},
  year      = {1993},
  month     = mar
}

\end{document}